# Emergent Growth of System Self-Organization & Self-Control

## Contextual system design, steering, and transformation

Author: Jessie Henshaw, HDS natural systems design science; research consultancy
680 Ft Washington Ave, New York NY

sy@synapse9.com

Abstract: In physics, I noticed subjects not explained by formulas were often not studied, like how uncontrolled growth systems changed form. Weather, businesses, societies, environments, communities, cultures, groups, relationships, lives, and livelihoods all do it following some variation of an 'S' curve. It is a slow-fast-slow process of self-animated contextual energy system development. Making working systems, they seem to develop by "find and connect" in three stages, starting small to first A) freely build designs of growing power, then B) diversify, adapt, respond, and harmonize with others, then C) take on one or more roles in their climax environments. It is a life curve of tremendous syntropic success that then ends with decline. Life is particularly risky for small start-ups, but many do succeed. Many powerful civilizations have emerged, some never growing up *but growing as endless startups,* only to become fragile, fail, and vanish. That looks like a choice.

Credit Author Statement:   Concepts and figures are by the author UON.

Sponsor Statement:         The work was self-directed and self-funded

## Practitioner Points: Contextual studies of system change

A. The steering and transformations of systems in context start during growth. Capturing energy and resources animates the expansion of the system building on itself. Understanding what one sees happening may require changing viewpoints as the process connects the parts and changes form.

B. System change alternates between speeding up and slowing down, requiring different kinds of accumulative organization for continuity. As an organizational rather than mathematical process, forming links has to work by both "find and connect" and "cause and effect.

C. Systems form internalized an organization as they emerge from their contexts, coupled intermittently with their resources by their internal working relationships separated from outside view and by only working as a whole. So, working independently, except as pathfinders for common interests, they then make wonderfully organized but blind worlds.

## I. Introduction

T hroughout our evolution, we learned from what was observable about the systems that mattered to us. We found languages to connect our meanings with what we noticed, saw, felt, and heard[1]. Growth gave us both our own lives and the wealth of nature, our good relationships, societies, and even the great conflicts and crises that beset us, all by continuities of cumulative design for building internal structures by capturing external resources observable from the outside but internally self-referential and so rather hidden from view. Ways of understanding these coupled external/external structures to make some sense of them are in the author's earlier writings (Author 1985, 1995, 1999, 2007, 2010, 2018, 2019, 2021, 2022) and, of course, also in everyone's successful experience of living with them. Here the focus is on studying the natural designs produced by growth in context, keeping the stiff requirements for energy continuity and conservation in mind.

The earliest term found for the emergence of complex systems by growth is the ancient Greek term *physis*.[2] That became the root of the new term *physics* as Greek science later became formalized, referring to all the deterministic processes of nature. That change seems to have been a demographic one, a branching of knowledge cultures going their separate ways. The original reference is to nature, and its creative design processes are still current in Greek. The newer, exceptionally powerful deterministic modern physics meaning, of course, now dominates the world. Its great economic power as a language then naturally paralleled the use of its tools that empowered economic growth, becoming a metaphor for knowledge in general.

There is a great hidden irony in how the profitability of science seemed to project limitless economic growth, partly in the making of powerful concepts detached from their contexts. It tends to hide what becomes of them when used. As growth works in nature by coupling systems with contexts, detaching the rules of science from their contexts would have been driven by the economic demands for maximizing profits. That started at the very dawn of science. The fame of the first Greek scientist, Thales, came from collecting all the mathematics of antiquity and making huge profits in trade from using it,[3] proving the great power of profiting from sound equations.

For Aristotle, Plato, and the scientific communities that followed, the great financial rewards for the profitable parts of science also enabled extraordinarily diverse kinds of learning of every kind. The concentrated essentials of the craft became what we now know as physics[4] It also still seems to be the unusual profits from abstract principles, detached from the contexts of their use, that drive today's economy. As that detachment of theory from context seems to be part of our problem, this work aims to help start a reconnection of our great abstract thinking with the natural systems it idealizes.

---

[1] Author 2022 talk on our first "systems science" "Language as a knowledge tree for systems in context" https://www.youtube.com/watch?v=Xd3aLWMztR0

[2] "physis" - ancient Greek term for "nature", from the verbal noun φύσις, "phusis", meaning "growing", "becoming", itself from φύω, "to grow", "to appear". https://www.etymonline.com/search?q=physis

[3] References to Thales' Science and Philosophy https://www.google.com/search?q=thales+science+philosophy

[4] Physics was known as natural philosophy until the late 18th century. By the 19th century, physics was realized as a discipline distinct from philosophy and the other sciences.. Wikipedia  bit.ly/3LRcd8B

Two of the scientists who offered the most useful observations on the creativity of growth were Ken Boulding (1953) and his teacher J M Keynes (1933), both of whom took a diagnostic approach similar to the one used here. The early field studies that led to this work were of how air currents developed (Author 1978). Of course, the wonderful observations of many others, Bogdonov, Bernalanffy, Ashby, Bateson, Troncale, Tainter, Klir, Miller, Varela, Maturana, Walter Elsasser, and Al Bartlet, the architects Christopher. Alexander & Louis Kahn, paleontologist Stephen Gould, and zoologist D'Arcy Thompson to name a few.

None but Keynes and Boulding seemed to notice the main question of how growth both animated and limited systems. That science focused on fixed rules in a world full of variously directed creative processes does seem to be what led to our basing our economic models on limitless growth. Casual observation tells one that growth is nature's most creative and self-limiting process, a contradiction of perceptions that even seems to go back millennia.

## THE NEW MODEL

In nature, growth is a pair of slow-fast-slow progressions, rising then declining. Five main stages of maturing self-organization may be visible too. Figure 1 labels them; A, B, C, D, & E. They're all part of growth. Today the term is used mostly for A, the takeoff period, such as for the economy, business startups, or the rapid formation of new lives in the egg or womb. It is also the initial conceptual stages of creative imagination and the infatuations leading to new relationships. If the startup lasts, the big question, there is much more to come.

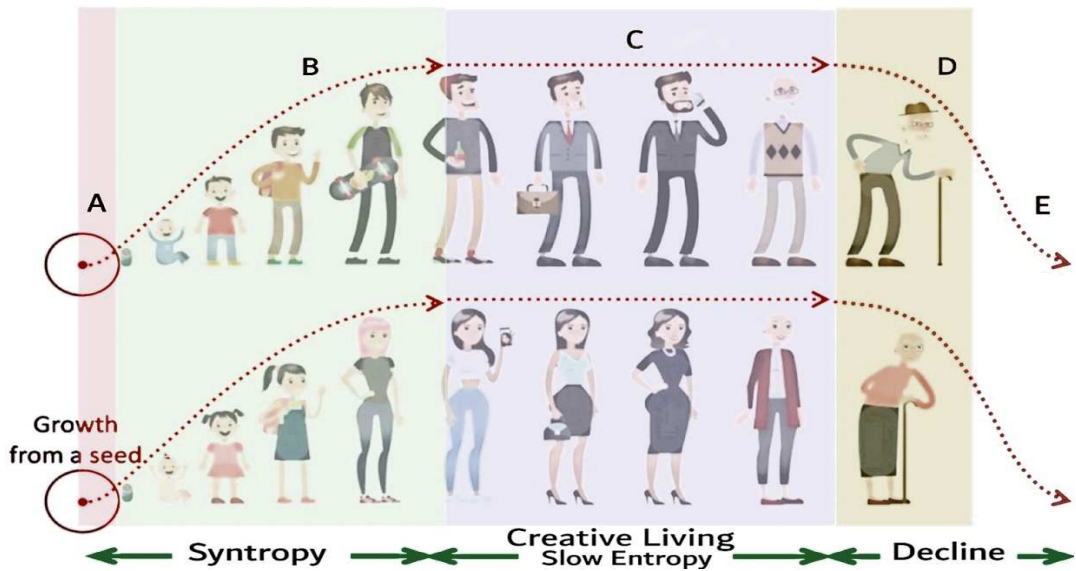

Figure 1. Our normal life cycle is a general model for things that grow: stage A) an explosion of new design expanding on a seed pattern in a protected place, B) a period of growing up, maturing the immature form, learning about its new place, C) a long life of creative engagement, D) a mostly comfortable detachment from others, then decline, E) weakening and approaching the end. (part credit - Getty Images)

### THE ISSUES

This approach is a bit jarring to scientists at first. It is about how uncontrolled systems work by themselves. The view of "everything is connected" differs too. It is quite true that relationships have surprising remote connections that may be very meaningful. Still, "everything is connected" seems untrue if you consider the major separations of life. Homes offer a model of how natural systems provide enclosures with connections. As for our bodies, homes, institutions, etc., the openings are all regulated. So some things become truly separated. That leads to gaps between some functions and information and the great value of enclosures allowing things in them to work differently.

Given humanity's tremendous capability of making things, our multiplying world crises[5] show humanity is missing something big. Our ever-growing crises appear to come mostly from blindly using our rules for profit to push our control of previously self-controlled contexts beyond their limits (Author, 2020). These confusing ironies are valuable "gaps" in our information, many of which may be real openings to what lies hidden. The tragedy is despite the past 50 years of great effort to solve our world crises they're only getting far worse (Meadows 1972, 2004; WEF 2023).

The life curve of growth shows that our crises seem to come from our failure to move from stage A to B. Normal growth systems respond to signals of growing collisions with other lives and internal distress. Responsive systems would move to stage B, giving them the time needed to mature and care for all their relationships and have a long, rich life.

Also ironic is the number of familiar examples to learn from. In a short paper on general concepts like this, exploring examples needs to be up to the reader, thinking over their experiences and daily life, using some of these ideas. Everyone has had multiple startup successes and failures, too. They are great examples of what to respond to in creating things that work. However, ironically designed to respond to profit-making, our world economy seems unresponsive to its survival. Why is that? It is led mostly by very well-educated, caring moms and dads, isn't it?

## II.   Materials and Methods

### THE PHYSICS

The physics for natural systems in context began with algorithms for reconstructing the continuities of time series data, to avoid using equations to fit data. Equations erase the natural dynamics and other evidence of the active systems. A spline curve is responsive to changing rates but in a more regulated way than nature. Figure 2, shows a derivative reconstruction (DR) for finding and verifying detailed shapes of a gamma-ray burst.[6] (Author 1995, 1999).

---

[5] An experimental list of World Crises Growing with Growth - https://synapse9.com/_r3ref/100CrisesTable.pdf  and Wikipedia https://bit.ly/3oSMeEG

[6] NASA report https://imagine.gsfc.nasa.gov/science/objects/bursts1.html

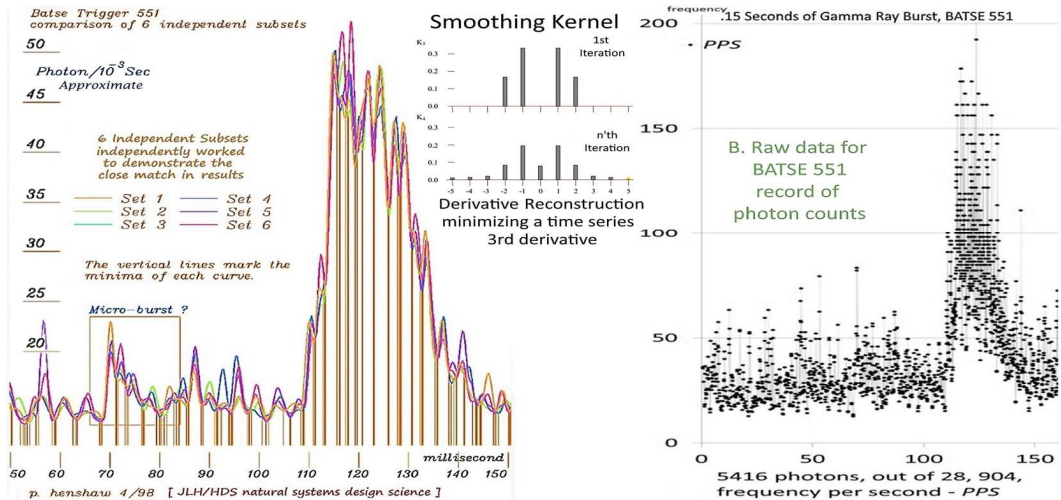

Figure 2. Dynamic details of the Gamma-Ray Burst BATSE-551: – Six divisions of the raw photomultiplier photon frequency data on the right into partitions of every 6$^{th}$ point, with derivative reconstruction for each, are on the left. The DR smoothing kernels (top center) are calculated to reduce the 3$^{rd}$ derivative at the center of five data point segments at a time for each series.

On the left in Figure 2 is the DR reconstructed continuities and on the right the sequential photomultiplier data. The validity of the reconstructed dynamics on the left varies with the coincidence of the shapes in the six subsets composed of every 6$^{th}$ point. The more coincident shapes indicate natural phenomena made more possible to study with this presentation of the data. The DR works by scanning each data set, repeatedly adjusting the continuity. The 3$^{rd}$ derivative (the 'jerk') is set to zero at the midpoint of each 5-point segment, to produce a "center-lightened" smoothing kernel with a hole in the middle, rather than a "center-weighted" one. Repeating ten times widens the effect, to produce each of the six separate curves compared.

## IMPLICATIONS OF CONTINUITY

Energy conservation implies continuity in all natural processes. A theorem and discussions (Author 1995, 2010) show that only assuming that change can't be instantaneous implies change must follow a gap-free developmental process. The scientific method teaches a search for the formula that processes fit, but not where the fixed formulas come from nor how the material processes that produce natural behaviors work. Understanding those processes is the focus here.

Process continuity changes the normal assumption from systems operating by their measures to operating by their processes. For continuous change, processes need flowing organization connecting their working parts, as seen in growth and elsewhere. Growth is a continuous transition between states with periods of regular proportional change as the change regularly doubles in scale. The process might be a flexible reproduction that multiplies, like cell or pattern replication, as seen in many kinds of systems that exhibit a variety of peak scales.

## A GUIDE FOR OBSERVATION

Figure 3 guides one's inquiries into how emerging systems develop internally organized and environmentally coupled relationships. Image depicts natural systems developing by a kind of flowing contagion of energy and organization concentrating processes, life working by itself, developing structures and connections.

Recognizing where trends of regular proportional change (regular feedback) start and end helps direct one's observations in the context to what is actively happening. Causation in natural systems also "has many parents," of course. A fertile environment is full of existing designs and resources that make foundations and leave openings for what animates the internal design rarely, if ever, directly seen, but as in life, leaving trails of different kinds recognized as signs or signals of what else is happening.

That certainly does not open the worlds of self-contained systems but exposes just enough for people paying attention to see what others are dealing with and when responses are needed. What we all know best are changes in response, pressures, stability, and regularity in anything, telling volumes about what is happening. It would matter tremendously if we recognized the connections between system transformations and the diagnostic signs of systemic needs, struggles, and vitality for the health of our environments and economies.

One then associates the shapes of change with the opportunistic feedback processes there are hints in the context, looking to distinguish any deterministic influences from the opportunistic processes that energize the system and let it make connections (Author 1985). How any person, community, relationship, or economy responds to opportunity will be characteristically different from one another and from one situation to another. What one can expect to be the same is that all the transitions will occur by some path of continuity.

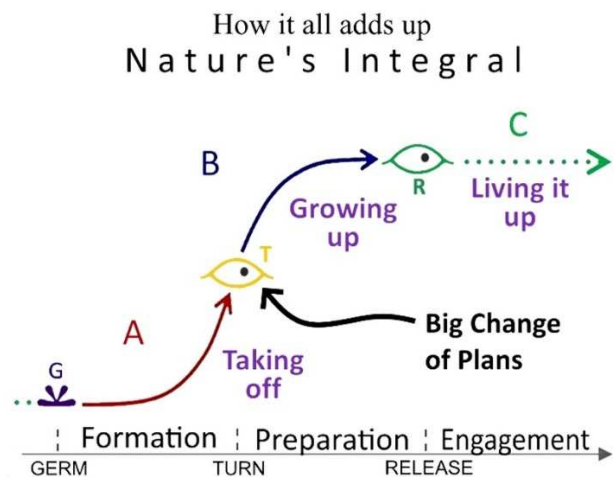

Figure 3. <u>The typical S curve stages of natural organizational change:</u> The stages of normal lives all begin with transformative events to start characteristic forms of organizational development, types: A) formation, B) preparation, & C) engagement; begun respectively by events: G-Germination, T-Turn-forward, C-Connect. The names suggest what to be ready

for or to look for evidence of. In common use, there are other names for them. Of course, later events of decline generally follow too.

It is not a focus here, but it is also useful to give attention to the attraction of opportunities and the action of taking them, relating to the Yin and Yang of receptive/assertive and passive/active relations. The view here is that those become quite complex sometimes, and the passive can often dominate, such as with advertising, pheromones, or special offers. So one should assume that initiating emergent changes might require both pushes and pulls. The best resource is either changes you can closely and repeatedly study or the familiar or special experiences one can recall graphically to repeatedly study and look for the kinds of situations and causations.

## III. Results

### CENTERS

Growth starts with an animating central design for capturing local resources to expand itself as a central culture of relationships forming an external world of connections. The most useful analogy is a home with a family coming and going to make their lives. We all know many kinds of families and understand neighborhoods, groups of friends, businesses, and nations as families too. "Centers with connections" is nature's main design.

The uniqueness of the cultures within homes comes partly from their independence. It may also come from each part incorporating the seed design of its origin, as in how a body's cells all contain the genome of the whole. Speakers of any language need images of how the whole works. Social and work groups also develop private languages, with each member carrying the group's unique way of connecting (Author 2022). These unifying features of whole systems hint at their possible genuine holographic design, giving them their holistic properties, seeming to imply active molecular orders of kinds to perhaps explore.

Startups without a protected place to start generally do not survive, though. So the first job of growth systems that will survive life encounters is to find more and different resources as the small resources of small things get used. Most survivors come from resourced and protected places they can freely colonize. The big change in plans comes when that free growth place and period runs out.

To survive, a radical change in direction comes from the pushes & pulls of holding the emergent life system together as its external resources dramatically change, such as at birth, upon graduation, or other radical changes of context. We see such "coming of age" points of readiness for new challenges all over, as in "calms before the great storms" and new species punctuating the equilibria from hidden places (Gould 2007).

### A PLANKTON SPECIATION CURVE

Figure 4 shows a roughly S-shaped curve of a plankton evolution, stable before and after, connected by dramatic fluctuations. In this case, it is an unbroken million-year event, the bumpy plankton developing from the smaller smooth one. One could photograph the original samples studied by Malmgren (1983) or collect new ones from the deep-sea cores to show the changes in great living detail with an animated film+

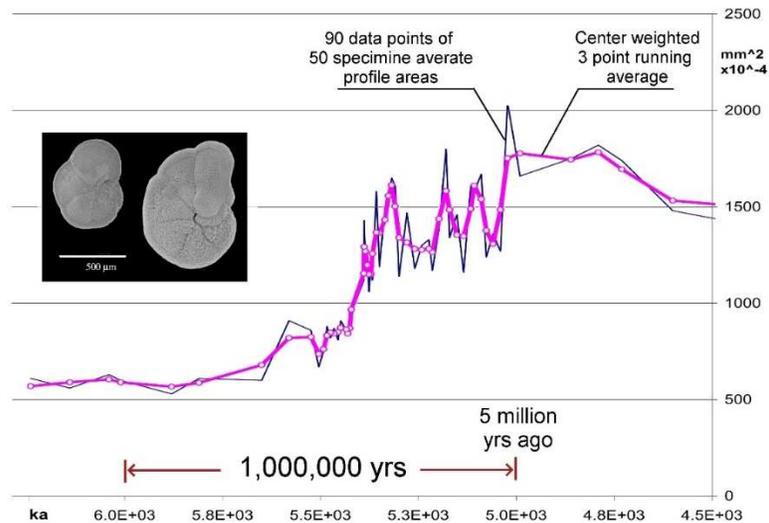

Figure 4. <u>Evolution of the small G. pleisotumida plankton to larger G. tumida:</u> (Malmgren et al. 1983; Author 2007) The evidence of large-scale continuity (rather than random walk) is from light smoothing reducing the double reversals (flip flops in direction) by ~75% for the G. tumida data. The strong effect shows noise to be local, not global, validating the light smoothing trends. For random series, fluctuations are little reduced. ─ Each of the 58 data points is an average profile area of about 50 from an Indian Ocean deep-sea sediment core.

The slow-fast-slow trend between levels of stability is clear, but what explains the remarkable irregularity? Successful smooth growth also starts by destabilizing and restoring a stable state with a new form. The extreme variation here hides periods of progression between the peaks, though. With that, the growth pattern is more like "trial and error," as very often seen in eventually successful struggles by people, businesses, social movements, and even weather as systems intermittently breakthrough (Author 2019), competitive struggles of the type being of the "try, try, again" variety.

This kind of "goal-directed struggle" seems at least remotely plausible here is a logical necessity for emerging systems to adaptively find resources to build onto its process for capturing more, and so need to have an exploratory ability to find more elsewhere, both as new system emerge, and as seen in how all mature species develop. Skillful search seems to be a general property of life.

## AN ECONOMY UNABLE TO RESPOND TO CHANGE

In Figure 5, Turn to the reader to ask, what do you see in the curves of the economy, the GDP-related impacts and resources, the 240 yr $CO_2$ growth curve, and the world eruption of refugees? Much has been said about humanity's amazingly creative destruction in using up the earth, but relatively little of why. Though life and growth are always risky, growth with an explosive start seems to be nature's way to avoid early hazards.

The homework is to ask why we're the third highly successful civilization in a row to head for self-destruction. There was the late Bronze Age east Mediterranean collapse of an early "united states of

prosperity[7]," Rome[8], and now us. You might read up on these and other civilization collapses to see what we're doing the same and differently and what the rules and exceptions are.

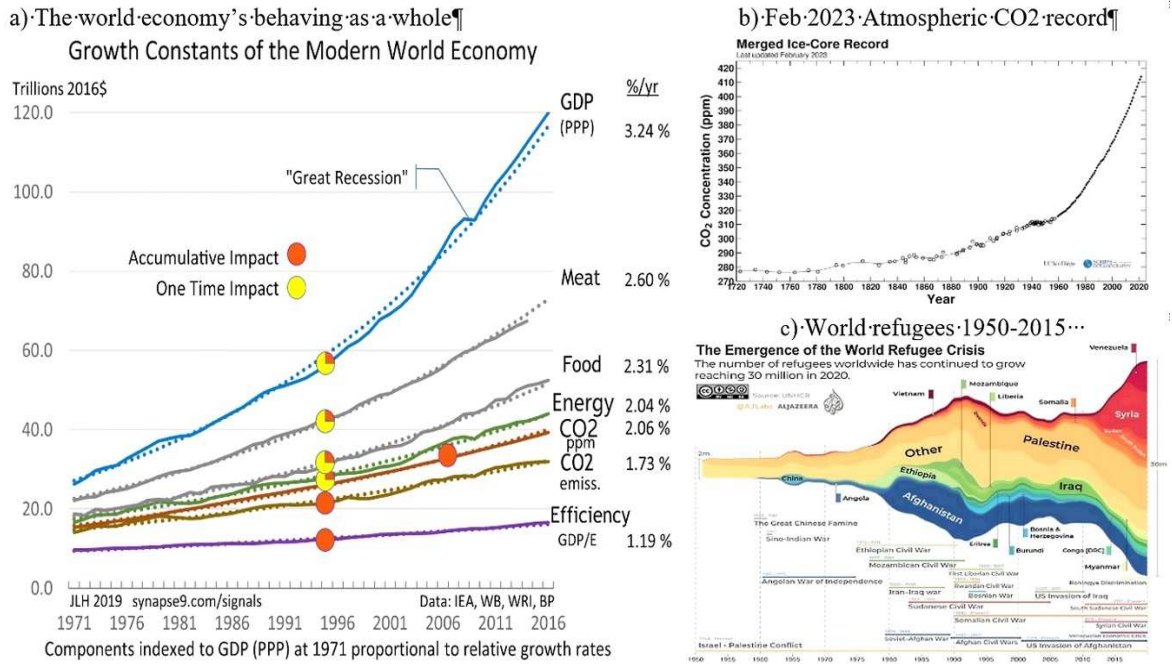

Figure 5. A Broad View of the State of the World: a)The constant growth coupling of world economic factors, b) the *unwavering* compound growth of atmospheric CO2 that drives climate change, c) the similarly explosive world refugee crisis.

### A CITY REGENERATED BY A CHANGE OF HEART

Interestingly nearly everyone who did something took full credit for the drug culture breaking like a feaver and disappearing like a bobsled on a ski mountain. The major bit of luck that added to the fragility of the mania was the emergence of Hip Hop in the middle of it. As the graffiti artists left the train yards, they did murals to their fallen friends, and the new culture and community options swept up all the kids in the Bronx who might have joined the drug world as it collapsed. So, in the end, the epidemic was overcome by the care and commitment of the whole city, which transformed it as a whole, too, from its long period of being a dangerous place to become fun, open, and free, remaining largely so now 30 years later, no wars! So it does seem horrible entrenched fixations can be changed if you work hard and have your feet on the ground.

---

[7] https://en.m.wikipedia.org/wiki/Late_Bronze_Age_collapse
[8] https://en.wikipedia.org/wiki/Ancient_Rome

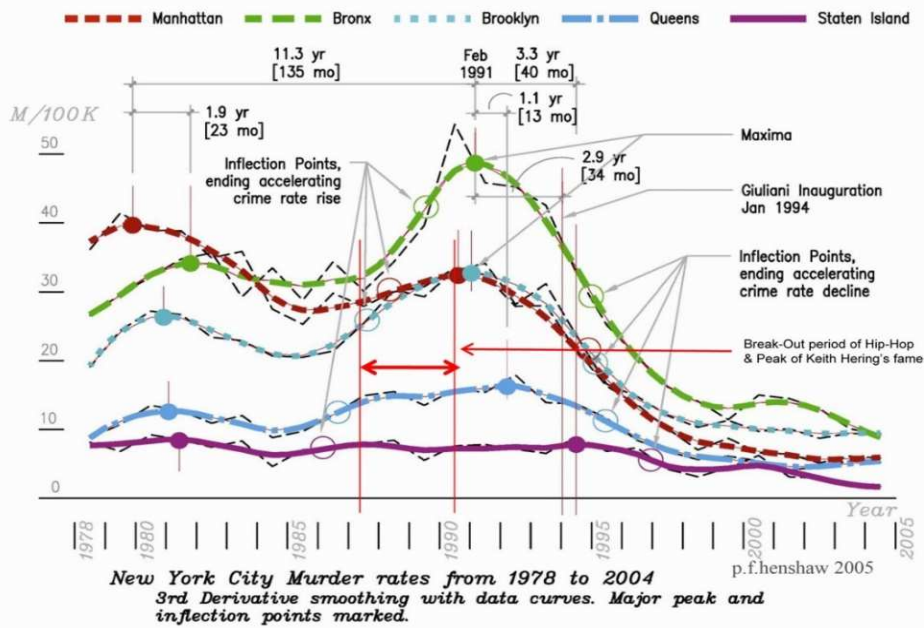

Figure 6. <u>Ending of the 40-year New York City drug & crime cultures wave:</u> The simple story is of an almost operatic drug dealing "gang banging" show-off gang crime wave that got the whole city to join in doing everything possible to end it, cured by love in the end it seems.

## IV. Discussion

### CONTEXT BLINDNESS

Understanding observation is important to a new science based on observing things in context. There is so much we can't know. Our long survival on earth means we know *something*, at least proving there *are* things to know. The key seems to be learning to appreciate the contexts. Contexts affect the meanings and effects of things, each locale a hidden world of local and remote relationships. A "newcomer" is blind to local customs, and so is a student entering a new year in school. We're also blind to the mental contexts of others and are often "out of the loop" among friends or colleagues. We're generally blind to how we're perceived and who others are. So we and our social groups are free to make up "tall tales" about ourselves and others. That we adopt these conflicting stories as real to us is very unreal, though, a double blindness.

Today's extremely misleading stories people make up about each other seem driven by a common source. They seem connected to the escalating struggle, pressures, and disruptions of everyone's lives we're experiencing. We're deservedly fearful and angry with no direct threat to point to. The apparent cause is our whole economic system pushing disruptive limits of both human and natural systems. Usually, responding to symptoms allows things to heal, but despite great efforts, they're not healing but still growing.

So we are a bit blind to how natural systems work. They develop by growing from the inside as our bodies did from a cell or as a classroom 'assembles'(creates its working form) in a room. That creates an exclusive home for its cultures, its internal working relationships the barrier to others participating. So their home relations are internalized and invisible from the outside. Along with that comes a universal blindness to how all systems work, which seems to be how nature works, and that we've been blind to.

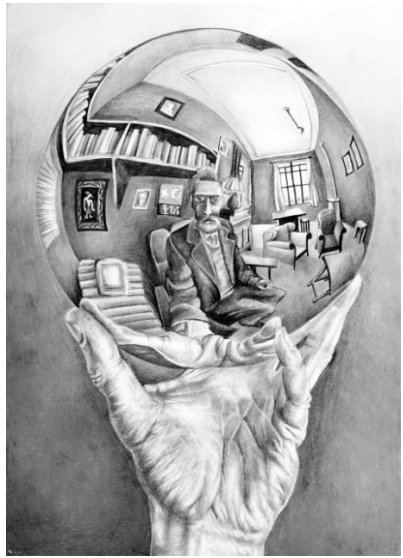

Figure 7. <u>Escher's Hand with Reflecting Sphere:</u> The irony expressed in the image seems to be that however you try to look at the insides of something else, all you often can find are reflections of one's self, as the interiors of nature are remarkably well hidden. That the effect may be due to something simple, like the internal relationships referring to themselves, shouldn't detract from the mysterious and amazing things they do.

That's not the end of the story, but an important start. When these confusing conflicts arise one on one, they're mostly curable by exposure to the contexts people are dealing with. The 2007 Nobel Prize in Economics (Ostrom 2009) was awarded for defining ways to introduce people to the world they need to care for. We now need major global efforts to learn that. There are varied other good methods too. Investing in removing our blinders is one of the things we can all do.

## V. Conclusion

Hopefully, this work will help others continue well-grounded studies of exploratory natural systems. The most difficult work involved was finding a comfortable language for discussing it. Repeated efforts to translate it for one or another community were unsuccessful. I persisted partly because the earliest versions from the 70s and the technical methods developed in the 90s held up. I still didn't find solid footing for it until I settled on working from the roots of common language, the meanings familiar to the wide community of languages that emerged from the Indo-European context. Those came from the Stone and Bronze Ages, a language only seen in the commonality of roots of all the many daughter languages that spread from there. That way of finally finding my footing was a tremendous relief!

## VI. Acknowledgments:

Growing up, I had freedom and wonderful guidance on questioning what I saw from my dad, an admired physics professor at Colgate Univ. My family descended from scientists and educators, too, surrounding me with people paying close attention and asking good questions. I also had amazing friends fascinated by unusual gaps, ironies, and misguided views to play with for entertainment. Going to a fine small-town school prepared me for the great learning available at my universities, St Lawrence, Columbia, and Pennsylvania. There were, of course, also the thrilling times, music, passions, travel, and experiments of the 60s and 70s that contributed as much as anything ─ pointing me right at the real problems ─ and quite a bit of luck.

## VII. References:

### A. Tables: none

### B. Data Accessibility & Figure Sources

#### 1. Figure 1 ─ Lifecycle Figures

Getty Images ID:906819280 - paid use 2081946114
https://www.istockphoto.com/vector/life-stages-set-gm906819280-249892188

#### 2. Figure 2 ─ Gamma-ray burst photon rates—

Data from NASA : https://synapse9.com/pubData/Ba551.txt

#### 3. Figure 3 ─ Nature's Integral

Digital drawing

#### 4. Figure 4 ─ G. tumida plankton dimensions —

Data from B. Malmgren: https://synapse9.com/pubData/Malmgren%20Data.xlsx

#### 5. Figure 5 ─ Economy Unable to Respond to Change – (triptic)

(a) Coupled World Economy Impacts

GDP (PPP) 1971 – 2016 IEA PPP data extended with recent World Bank data:
https://data.worldbank.org/indicator/NY.GDP.MKTP.PP.CD?end=2016&start=1990

World economic energy use 1965-2017 from BP:
https://www.bp.com/en/global/corporate/energy-economics/statistical-review-of-world-energy/downloads.html

Modern $CO_2$ Emissions – 1971-2016: IEA $CO_2$ data & WRI $CO_2$ emissions:
https://www.wri.org/resources/data-sets/cait-historical-emissions-data-countries-us-states-unfccc

Historical $Co_2$ Emissions 1751-2013: US DOE DOE CDIAC data: https://cdiac.ess-dive.lbl.gov/ftp/ndp030/global.1751_2014

World Meat Production – 1961-2016: Rosner - OurWorldInData:
https://ourworldindata.org/meat-and-seafood-production-consumption

World Food Production – 1961-2016: FAO: http://www.fao.org/faostat/en/#data/QI

(b) Figure 5b

Atmospheric CO2 Record: Scripps combined ice core CO2 ppm to1958, and average Mauna Loa and Antarctica mountain top from 1958. http://scrippsco2.ucsd.edu/data/atmospheric_co2/icecore_merged_products

(c) Figure 5c

World Refugee data: Al Jazeera: https://www.aljazeera.com/news/longform/2022/6/16/visualising-the-fastest-growing-refugee-crises-around-the-world   UNHCR data from: https://www.unhcr.org/global-trends

6.  **Figure 6 ─ A City Regenerated by A Change Of Heart**

NYC murder rates by borough from NY State murder rates by county. https://synapse9.com/pubData/NYS-murder_county.xls

## VIII. References

### A. Texts

Author  (1985). Directed opportunity, directed impetus: new tools for investigating autonomous causation. *Proceedings of the Society for General Systems Research*. Intersystems Publications. author copy: http://www.synapse9.com/pub/1985_DirOpp.pdf

Author  (1995). Reconstructing the physical continuity of events. An author research report on analytical methods developed. https://synapse9.com/pub/ContPrinciple95.pdf

Author (1999). Features of derivative continuity in shape, International Journal of Pattern Recognition and Artificial Intelligence (IJPRAI link to article), for a special issue on invariants in pattern recognition, V13 No 8 1999 1181-1199. Online - http://www. synapse9. com/fdcs-ph99-1. pdf

Author (2007). Flowing processes in a punctuated species change, G. *pleisotumida* to G. *tumida*, displaying feedback-driven evolution, pending resubmission. https://www.synapse9.com/pub/GTRevis-2007.pdf

Author (2010). The energy physics of continuity in change – draft pending revision. https://www.synapse9.com/pub/2010_drtheo.pdf

Author (2018). Systems-thinking for systems-making: Joining systems thought and action. *Systemic Practice and Action Research*, 32(1), 63–91. https://doi.org/10.1007/s11213-018-9450-2

Author (2019). Growth Constant Fingerprints of Economically Driven Climate Change: From 1780 origin to post-WWII great acceleration. draft pending revision. https://synapse9.com/drafts/2019_12-GrowthConstFingerprintsOfCC-preprint.pdf

Author (2020). Top 100 world crises growing with growth. Author experimental list. https://www.synapse9.com/_r3ref/100CrisesTable.pdf Accessed 04/13/2021

Author (2021). Understanding Nature's Purpose in Starting all New Lives with Compound Growth - New Science for Individual Systems. *ISSS 2021 Proceedings*. https://journals.isss.org/index.php/jisss/article/view/3911

Author (2022). Holistic Natural Systems - Design & Steering: Guiding New Science for Transformation. *International Society for the Systems Sciences* July 2022 Jul 9.

Boulding, K. E. (1953). Toward a general theory of growth. *The Canadian Journal of Economics and Political Science / Revue Canadienne d'Economique et de Science Politique*, 19(3), 326–340.  https://www.jstor.org/stable/138345?seq=1

Gould, S. J. (2007). *Punctuated equilibrium*. Harvard University Press. Grossman, K., & Paulette, T. (2020). Wealth-on-the-hoof and the low-power state: Caprines as capital in early Mesopotamia. *Journal of Anthropological Archaeology*,60, 101207. https://doi.org/10.1016/j.jaa.2020.101207

Keynes, J. M. (1935). The general theory of employment, interest, and money. Harcourt Brace Jovanovich. http://synapse9.com/ref/Keynes-ebook-TheGeneralTheory.pdf Accessed

Malmgren, B. A., W. A. Berggren, and G. P. Lohmann. (1983). Evidence for punctuated gradualism in the Late Neogene Globorotalia tumida lineage of planktonic foraminifera. Paleobiology 9:377-389

Meadows, D. H., Meadows, D. L., Randers, J., & Behrens, W. W. (1972). The limits to growth: A report for the Club of Rome's project on the predicament of mankind. A Potomac Associates Book.

Meadows, D. H., Randers, J., & Meadows, D. L. (2004). Limits to growth: The 30-year update. Chelsea Green Publishing.

Midgley, G., & Richardson, K. A. (2007). Systems Thinking for Community Involvement in Policy Analysis. *Emergence: Complexity & Organization*, 9. https://www.academia.edu/download/39922639/Systems_thinking_for_community_involveme20151112-10954-1nfvm3k.pdf

Ostrom, E. (1990). *Governing the commons: The evolution of institutions for collective action*. Cambridge University Press.

Ostrom, E. (2009). A general framework for analyzing sustainability of social-ecological systems. *Science*, 325(5939), 419–422. https://doi.org/10.1126/science.1172133

Ostrom, E. (2010). Beyond markets and states: polycentric governance of complex economic systems.*American economic review*,100(3), 641-72. https://doi.org/10.1257/aer.100.3.641

Steffen, W., Broadgate, W., Deutsch, L., Gaffney, O., & Ludwig, C. (2015). The trajectory of the Anthropocene: The great acceleration. *The Anthropocene Review*, 2(1), 81–98. https://doi.org/10.1177/2053019614564785

Tainter, J. (1988). *The collapse of complex societies*. Cambridge University Press.

WEF (2023) The Global Risks Report 2023. World Economic Forum https://www.weforum.org/reports/global-risks-report-2023/